\documentstyle[psfig,prb,twocolumn,aps]{revtex}

\begin{document}

\begin{widetext}

\title{Ab-Initio Calculation of the Metal-Insulator Transition in Sodium rings and chains
and in mixed Sodium-Lithium systems}

\author{Walter Alsheimer and Beate Paulus}

\address{ Max-Planck-Institut  f\"ur Physik komplexer Systeme,
N\"othnitzer Stra\ss e 38, D-01187 Dresden, Germany}

\maketitle

\begin{abstract}
We study how the Mott metal-insulator transition (MIT) 
is influenced when we deal with electrons with different angular momenta.
For lithium we found an essential effect when we include $p$-orbitals in the description of the 
Hilbert space. 
We apply quantum-chemical methods to sodium rings and chains in order
to investigate the analogue of a MIT, and how it is influenced by periodic and open boundaries. 
By changing the interatomic distance 
we analyse the character of the many-body wavefunction and the charge gap.
In the second part we mimic a behaviour found in the ionic Hubbard model, 
where a transition from a band to a Mott insulator occurs. For that purpose we perform
calculations for mixed sodium-lithium rings. 
In addition, we examine the question of bond alternation for the pure sodium system and the mixed sodium-lithium 
system, in order to determine  
under which conditions a Peierls distortion occurs.
\end{abstract}
\vspace*{2cm}
Pacs-No.: 71.30.h, 71.10.Fd, 31.25.Qm

\end{widetext}
\section{Introduction}

The metal-insulator transition is relatively well understood in the picture of
the Hubbard model (for an overview and references see, e.g., [\onlinecite{gebhard97}]).
There, the transition is driven by the ratio of the on-site Coulomb interaction $U$ and the hopping
term $t$. 
However in realistic systems it is normally insufficient to model the electronic structure
with a single hopping term and an on-site Coulomb term. Furthermore 
the question arises how orbitals with different angular momentum quantum numbers
influence the MIT. In the single-band Hubbard model only one $s$-type orbital per site is
supplied. 
It was found[\onlinecite{paulus03b}] using quantum chemical ab-initio methods 
that for lithium rings the inclusion of orbitals with $p$ character is 
essential to describe quantitatively the MIT in this system. If the $p$ orbitals are neglected,
the MIT occurs at a different position in the parameter space and the energy gap in the insulating regime is
much smaller. We model the ratio $\frac{U}{t}$ via the interatomic distance, where an increasing distance 
leads to a decreasing hopping. The on-site $U$ stays constant. 
The MIT occurs in a region of interatomic distances 
where the many-body wavefunction changes its character rapidly from
significant $p$ contribution to purely $s$ contribution.\\
In the present publication we analyse the MIT in pure sodium systems and a mixed sodium-lithium system.
The latter is a ring with alternating Na and Li atoms, chosen to mimic the situation encountered in the 
ionic Hubbard model\cite{fabrizio99,anusooya01,manmana03}.
There we have, for small
interatomic distances, a band insulator. When  increasing the interatomic distance sufficiently
a transition to a Mott insulator occurs. 
In addition we 
extend our studies from periodic boundary condition (Na rings) to
open boundary conditions (Na chains). The charge gap (ionisation potential minus electron affinity)
and the static electric dipole polarisability are calculated as a function of the interatomic distance.
The polarisability can be used as a measure for the MIT as pointed out by Resta and Sorella\cite{resta99}.
Furthermore the question of a possible Peierls distortion\cite{peierl55} is addressed. An insulator can also be formed
due to the electron-phonon coupling, where a lattice distortion yields a lower ground
state energy than equidistant arrangement of the atoms.\\
The paper is organised as follows: In Sec. II we present some technical details and
results for the Na$_2$ and the NaLi dimer. In Sec. III we discuss the influence of the boundary 
conditions in the pure Na systems and in Sec. IV we examine the mixed NaLi system.
Conclusions follow in Sec. V.

\section{Technical details}
\subsection{Basis sets}

Mostly Gaussian type basis sets are used in quantum chemical ab-initio methods  
to model the Hilbert space. For lithium a contracted $[4s1p]$ basis was used\cite{paulus03b}, that
was sufficient to describe the main feature of the system (quasi-degenerate $s$ and $p$ orbitals,
negatively charged ions) without extreme computational costs. The selection of a proper basis set
for sodium and the question of whether the use of a large-core pseudopotential is sensible for describing the core electrons
is addressed in this section.\\
All calculations are performed with the program package
MOLPRO\cite{molpro2002,wk1,wk2}.  
Starting from a mean-field Hartree-Fock (HF) description, we reoptimise the valence wavefunction in
multi-configuration self-consistent-field (MCSCF) calculations, keeping for most of the computations 
the $1s^22s^2p^6$ core electrons frozen at the HF level, if not indicated otherwise.
In this way the number of valence active-space orbitals is chosen to be equal to (or larger than) the number of Na atoms of the system,
and all possible occupancies of these orbitals are accounted for.
With this choice, we can properly describe the dissociation limit where each atom has one valence electron.
On top of the MCSCF calculations, we apply the multi-reference averaged coupled pair functional (MRACPF) method \cite{gdanitz88,ACPF} to deal with the dynamical correlations.
Here, all configuration-state functions are included which can be generated from the MCSCF reference wavefunction by means
of single and double excitations from the active orbitals.\\
We calculate the ionisation potential (IP), the electron affinity (EA) 
and the dipole polarisability of the Na atom. In all calculations
the core-valence correlations are neglected. We use a 10-electron pseudopotential\cite{fuentealba82} for Na
and the corresponding basis set is derived as follows. A contracted $[2s2p]$ basis is supplied with
the pseudopotential\cite{fuentealba82}. An extra diffuse $s$-function is added (exponent 0.0094) resulting
in the $[3s2p]$ basis. For the final $[3s1p]$ basis we neglect the second diffuse $p$ function and
recalculate the contraction coefficients for the remaining $p$ function from the $^2$P state at HF level.
The IP agrees well with experiment\cite{IP} (error $\approx$ 3.7\%).
With our basis we obtain 88\% of the experimental EA\cite{IP}.
The polarisability is overestimated by about 15\%, however it is known [\onlinecite{mueller83}]
that for this property the core-valence correlations
are important. 
Although our basis is relatively small we can nevertheless describe 
all quantities we are interested in with sufficient accuracy.

\subsection{Na$_2$ and NaLi dimers}

\begin{table}
\begin{tabular}{l|c|c|c|c|c}
&basis&$d_{\rm dimer}$&$\alpha_{xx}$&$\alpha_{zz}$&$\mu$\\
\hline
Na$_2$&ecp[3s1p]&3.392&219.5&469.5&\\
&ecp[3s2p]&3.310&211.6&449.2&\\
\hline
&exp\cite{huber79}&3.08&\multicolumn{2}{c|}{$\alpha_{\rm mol}=270$}&\\
&CISD(22$e$)\cite{antoine99}&3.09&207.8&360.4&\\
\hline
NaLi&ecp[3s1p],[4s1p]&3.106&180.3&387.0&0.649\\
&ecp[3s2p],[4s2p]&3.036& 181.1&395.1&0.565\\
\hline
&exp\cite{engelke82}&2.89&\multicolumn{2}{c|}{$\alpha_{\rm mol}=263$}&0.193\\
&CISD(14$e$)\cite{antoine99}&2.90&189.0&325.6&0.189\\
\end{tabular}
\caption{\label{dimer}
The equilibrium distance $d_{\rm dimer}$ in \AA ,the static dipole polarisability
per molecule perpendicular to the bond $\alpha_{xx}$ and in bond direction
$\alpha_{zz}$ in a.u. and the permanent dipole moment $\mu$ in a.u. are listed for
Na$_2$ and a NaLi dimer. The static dipole polarisability and permanent dipole moment
are calculated at the experimental dimer distance. The values are determined for different
basis sets and compared with experiment and literature.
}
\end{table}

For the dimers we perform MCSCF calculations with an active space of 4 orbitals (the occupation number of the
corresponding natural orbitals is larger than 0.2)
with the atomic closed shells kept frozen. On top of this MCSCF calculation
a MRACPF calculation provides the dynamical correlation. The results for the Na$_2$ dimer
and the NaLi dimer are listed in Table \ref{dimer}. 
With our selected basis set ecp[3s1p] for Na and [4s1p] for Li the dimer bond lengths are
significantly too large. Both additional $p$ functions and a $d$ polarisation function (data not in the table)
reduce the dimer equilibrium distance by about 0.1\AA . Core-valence correlations,
treated at the MRACPF level with the same basis as for the atom, reduce the dimer bond length
further, close to the experimental value\cite{huber79,engelke82}. The static dipole polarisability 
is not strongly dependent on the basis set used. Additional polarisation functions
reduce it slightly, by about 10\%, and the core-valence correlations have an influence
below 3\% within the basis we applied. But the core-valence correlations have a large influence on the permanent
dipole moment, where they reduce the dipole moment by more than
a factor of 2. When comparing with experiment, for Na$_2$ the polarisability is overestimated by only 12\%
(even for the smallest basis applied), for the NaLi dimer 
the best basis set applied including core-valence correlations underestimates the
experimental value by 9\%. The same was found by Antoine et. al.\cite{antoine99}
with CISD calculations.
Although the dimer data are not fully satisfying (for better agreement
with experiment we have to increase the basis set further and discuss different
correlations methods) we can show that our selected basis set is sufficient to
describe the main properties of the dimers and therefore of bound systems
which we want to analyse in the following sections. It is not our purpose to produce 
very good dimer data.

\section{ The MIT in the pure Na system with different boundary conditions}

\subsection{Na$_{10}$ ring and Na$_{10}$ chain at the equilibrium distance}

To compare open and periodic boundary conditions we have selected the equidistant Na$_{10}$ chain
and the equidistant Na$_{10}$ ring. As reference interatomic distance $a_0$ we have chosen
the Na-Na distance from the 3-dimensional crystal (3.659 \AA )\cite{kittel}. We calculate
the equilibrium interatomic distance and the cohesive energy per atom with different 
quantum chemical methods. At the closed shell Hartree-Fock (HF)
level, both the ring and the chain are not bound, the cohesive energy is positive. 
The equilibrium lattice constant
at the HF level is about 5\% smaller than $a_0$. A density-functional treatment with a LDA functional\cite{LDA}
yields the same lattice constant as HF, but the systems are bound. The LDA method works quite
well for the equilibrium ground-state properties, but cannot describe the 
dissociation limit in the systems. For that purpose a multi-reference treatment is necessary
with at least as many active orbitals as there are atoms in the system. We perform a
MCSCF calculation with 11 active orbitals both for the Na$_{10}$ ring and chain.  
On top of it a MRACPF calculation with single and double excitations using the same active space is applied.
The equilibrium distance does not change much.
The cohesive energy for the ring is 2/3 due to static correlations (MCSCF) and 1/3 due to dynamical
correlations (MRACPF), whereas for the chain the dynamical correlation provides more than half of the cohesive energy.
Comparing the MRACPF cohesive energy with the LDA one, we notice that LDA  
slightly over-binds the system. Increasing the basis set should bring the MRACPF and LDA values closer to each other.\\
When comparing the ring and the chain, we notice that both have nearly the same lattice constant, but the binding in
the chain is about 20\% weaker. The cohesive energy of the Na$_2$ dimer with the same basis set is 
0.0077 a.u. per atom and therefore 23\% weaker than the chain and 35\% weaker than the ring. Both can be regarded
as a system of 5 dimers.
Overall, the ring and the chain system have quite similar equilibrium ground-state properties.
The boundary conditions have nearly no influence when the atoms remain equidistant.

\subsection{The characteristic features  of the ground-state wavefunction}

\begin{figure}
\begin{center}
\psfig{figure=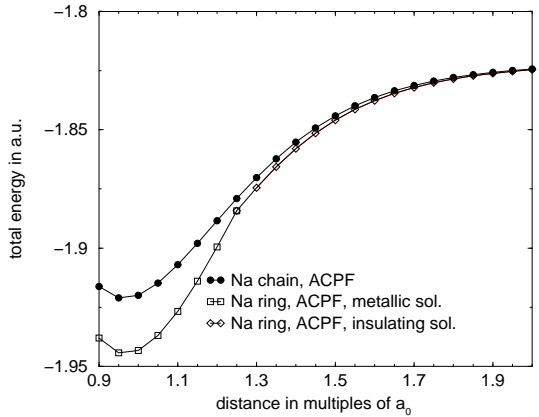,angle=-90,width=7cm}
\end{center}
\caption{\label{nawf} MRACPF energies of the Na$_{10}$ ring and the Na$_{10}$ chain versus the 
Na---Na distance. The curves labelled  "insulating solution" and "metallic solution" refer to different
MCSCF zeroth-order wavefunctions used as reference for the subsequent MRACPF, cf.\ text.}
\end{figure}

\begin{figure}
\begin{center}
\psfig{figure=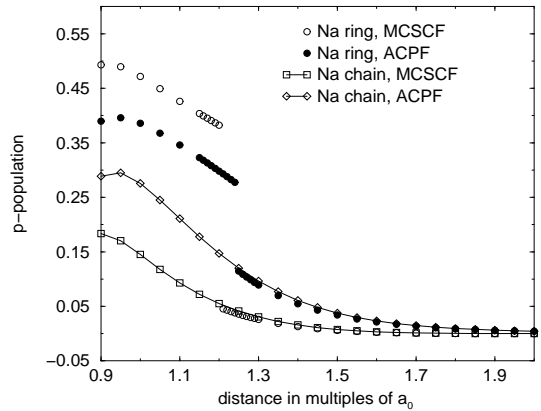,angle=-90,width=7cm}
\end{center}
\caption{\label{napocc} The $p$ occupation of the Na atom in a Na$_{10}$ ring
and of the inner Na atom in a  Na$_{10}$ chain 
versus the Na---Na distance. The $p$ occupations are calculated using the Mulliken 
population analysis.} 
\end{figure}

We want to investigate how the character of the many-body ground-state
wavefunction changes when enlarging the interatomic distance from
$a=0.9a_0$ to $a=2.0a_0$. 
For that purpose we perform a MCSCF calculation\cite{MCSCF1,MCSCF2} for the lowest singlet state,
on top of a closed-shell
HF calculation.
The selection of the active space
is crucial:
For the Na$_{10}$ ring we found (as for the Li system)
that not all important orbitals for the equilibrium distance and the
dissociation limit (10 important orbitals) fall into the
same irreducible representation. One of the important orbitals
at the equilibrium distance falls into a different representation than the 10 orbitals of the dissociated limit.
Therefore, we select as common active space the union of 11 orbitals.  
These orbitals were reoptimised at the MCSCF level. Finally, we performed a MRACPF calculation\cite{gdanitz88,ACPF} on top
of the MCSCF, in order to include dynamical correlation effects as completely as possible.\\
For the Na$_{10}$ chain we found that the 10 important orbitals of the dissociated state
are in the same irreducible representations as the 10 important orbitals at the equilibrium
distance. This is different to the ring system with periodic boundary conditions.
Therefore we perform here a MCSCF calculation with 10 active orbitals
and proceed with a MRACPF calculation as for the ring.
For the Na$_{10}$ chain it is interesting to look at the net charge and the
$p$ contribution of the charge population for the single atoms. 
Near the equilibrium lattice constant the $p$ population is high ($\approx$ 0.25) and quite constant
for the 6 inner atoms, and drops off for atoms belonging to the chain end. 
Concerning the charge transfer to the open boundaries, only the two inner atoms
are almost neutral, whereas at the boundary some oscillations smaller than 0.05$e$ occur.
For an interatomic distance near the dissociation limit, the oscillations in the
net charge disappear, all atoms are neutral. The $p$ population is constant and quite small ($\le$ 0.04) for all
atoms.\\
Analysing the ground state energy for the Na ring we found as for the Li ring 
a kink in the total energy curve.
Although we included in the active space all orbitals which are important in 
the limits of small and large atomic distances,
the total energy
as a function of the Na---Na distance is still not a smooth curve (Fig. \ref{nawf}),
because the nature of the active orbitals changes along the curve.
There is a curious hysteresis-like behaviour, namely
when starting the calculations from large $a$ (wavefunction for the insulating state) and always using
the previous solution as a starting point for the next smaller lattice constant, this yields
a slightly different solution in the region of the MIT than when starting from the metallic regime and
increasing $a$.
For the linear Na$_{10}$ chain we found a smooth curve, which coincides with
the ring for large interatomic distances, but is higher in energy
for the equilibrium distance.
To analyse the kink in the ring system in more detail, we perform a Mulliken population analysis for the $p$ orbitals 
(Fig. \ref{napocc}) of the MRACPF wavefunction as well as of the underlying MCSCF wavefunction
as a
function of $a$.
Whereas for the chain the $p$ population increases smoothly from the dissociation limit,
in the ring system we found a jump to significantly higher populations at 1.25$a_0$ at the MRACPF level and
at 1.21$a_0$ at the MCSCF level.
If we would characterise the metallicity by using the $p$ population analysis, we should pin the MIT in
the region where the jump of the $p$ population occurs.
But that would mean that the chain system has no metallic character and only the periodic
boundary conditions force the system to be metallic.
Therefore we look in the following section at the one-particle energy gap and compare for 
this quantity the ring and the chain system.

\subsection{The one-particle energy gap}

\begin{figure}
\begin{center}
\psfig{figure=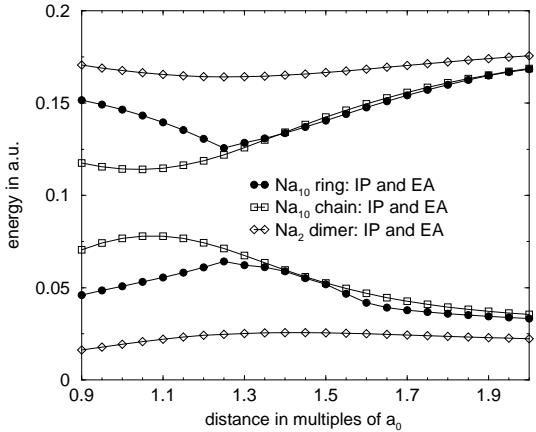,angle=-90,width=7cm}
\end{center}
\caption{\label{nagap} MRACPF values for the EA and IP
of the Na$_{10}$ ring, the Na$_{10}$ chain and the Na$_2$ dimer 
versus the Na---Na distance.
}
\end{figure}

The energy gap is determined by the energy difference between the ground state of
the neutral system and the ground state of the systems with one electron added and one subtracted.
We
calculate the MRACPF ground-state energies of the Na$_n^-$, Na$_n$ and Na$_n^+$ systems
and determine  EA=E$({\rm Na}_n)$-E(${\rm Na}_n^-)$ and  IP=E(Na$_n^+$)-E(Na$_n$), and
from IP-EA the corresponding gap. 
For a true metallic solid adding and removing an electron 
costs an energy given by the chemical potential. 
In finite systems, there always remains the difference between the one-particle energies
of the additional and the missing electron as well as the influence of relaxation and correlation effects.\\
For the Li$_n$ rings a finite-size analysis was performed\cite{paulus03b}. The gap energy monotonously decreases
with increasing number of atoms in the ring; a linear decrease is found up to $n=12$, the largest system
that is possible to treat with MRACPF. For all investigated rings ($n=2,6,8,10,12$)
we found similar behaviour with increasing distance independent of the number of atoms in the ring.
In spite of this we reemphasise that,
although we have a finite system instead of a real metallic system,
we still see a transition from a metallic-like regime
(where the gap is closing with increasing Li-Li distance)
to an insulating regime where the gap is
opening towards the atomic limit.\\ 
The EA and IP for the Na$_{10}$ ring and chain, and for comparison also for 
the Na$_2$ dimer, are plotted in Fig. \ref{nagap}.
For the dissociation limit
where HF fails to describe correctly the dissociation, the MRACPF calculation  yields the correct atomic IP and EA
for all systems evaluated.
The differences occur when decreasing the interatomic distance. For the Na$_2$ dimer the gap is closing
very slowly and below a region of 1.4$a_0$ it opens again slightly.
For the Na$_{10}$ chain the gap is also closing smoothly, but much faster than for the dimer.
There the minimal gap occurs at about 1.05$a_0$. In the insulating regime
the ring and the chain system behave very similarly, but when decreasing the 
interatomic distance the closing of the gap ends suddenly for the ring at that point, where the
character of the wavefunction changes. Starting from there, the gap opens again
and behaves like  
a free electron system in a box when decreasing the box length.
This opening of the gap would not be so pronounced if we could perform computations in a
ring with more than 10 atoms.\\
The behaviour in the insulating regime is independent of boundary conditions.
At the equilibrium distance (around $a_0$) the gap for the chain is smaller by a factor
of 2 than in the ring system.
It is not easy to define  a point for the linear system, where the MIT occurs,
but taking the minimal gap as an indicator for the MIT transition, the
metallic like behaviour develops for the chain at much smaller distances than for the ring.
Overall the boundary conditions do not change the qualitative picture,
but have a large influence on the quantitative one.

\subsection{Lattice distortion}

In this section we still study the pure Na system, but we allow for a bond alternation along the ring and the chain, 
so that the Na atoms can dimerise.
For average lattice constants larger than that where the metal-insulator transition occurs 
(i.e., for lattice constants where the insulating
solution is the ground state) the formation of dimers is energetically favoured, both for the ring and the chain system. 
The energy gain is due 
to bond formation between two Na atoms. For the ring in the metallic regime ($a\le a_{\rm MIT}$) 
a Peierls distortion\cite{peierl55} occurs, stabilising
the dimerised state. 
However, decreasing the lattice constant further, we find a point ($a=1.1 a_0$ for Na$_{10}$ ring)
below which the equidistant arrangement is 
the ground state, which would imply a metallic state for the infinite chain. This finding is analogous to the
one reported for the Li rings\cite{paulus03b}.\\
For the chain system we have a different result. There a dimerisation occurs for all
mean Na---Na distances. Even for a mean distance smaller than the
distance of the free Na---Na dimer with the same basis set (0.9$a_0$), the systems
show a slight bond alternation. This is due to the open boundary conditions,
which favour bond alternation. In comparison to the periodic boundary conditions, where
an equidistant arrangement is favoured for the equilibrium interatomic distance, the system with open boundaries
always shows dimerisation. 
Here the boundary conditions even have a qualitative effect on the system, and only for
infinite linear chains would the results coincide with the ones of periodic boundary conditions.

\section{Mixed NaLi ring - a realisation of the ionic Hubbard model} 

\subsection{ Hartree-Fock band structure}

\begin{figure}
\begin{center}
\psfig{figure=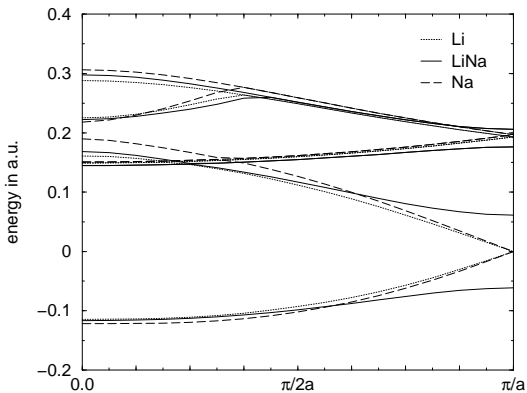,angle=-90,width=7cm}
\end{center}
\caption{\label{linaband} Hartree-Fock band structure for the infinite NaLi chain
and for the Li and Na chain back-folded to the NaLi Brillouin zone.
All systems calculated at the same lattice constant (interatomic distance = 5.954 a.u.).
The Fermi energy is set to zero.}
\end{figure}

As a first step we study the HF band structure of the one-dimensional infinite NaLi system
and compare with that of the pure Li and Na system. 
Using the Crystal program\cite{crystal98} we perform HF calculations for one-dimensional infinite chains.
As a basis set in Crystal we select for Li the optimised [4s3p1d] basis set of the three-dimensional metal\cite{doll99}.
For Na we choose the [4s3p] basis from Dovesi et al.\cite{dovesi91} and add a diffuse $sp$-function with exponent 0.08
and a $d$-function with exponent 0.4. 
In Fig.\ref{linaband} the Hartree-Fock (HF) band structure is shown for the optimised interatomic Li-Na distance of
5.954 a.u.. In addition we plotted the back-folded Li and Na band-structure at the same interatomic distances.
The Fermi energy is always shifted for all systems to zero. The main difference between NaLi and pure Na or Li 
occurs at the edge of the 
Brillouin zone, where the pure Na or Li system has no gap, whereas the mixed NaLi system
has a gap of about 0.12 a.u.. This clearly shows that NaLi is a band insulator, whereas Li and Na are metals
at the HF level.
The $s$ and $p$ mixing around the centre of the Brillouin zone is about the same for all systems. 
The population analysis in the mixed system yield 3.41$e$ for Li and
10.59$e$ for Na, i.e. nearly half an electron moves from the Na atom to the Li atom.
At the HF level the equidistant arrangement of the Li and Na atoms is stable, no dimerisation occurs in contrast to 
the pure Na or Li systems, whereas at the HF level we found a Peierls distortion.\\
An enlargement of the interatomic distance would not change the HF band structure significantly.
The $s$ and $p$ mixing will be reduced, but the LiNa system will remain a band insulator, as
Li and Na remain metallic. For an interatomic distance equal to twice the equilibrium distance
the charge transfer from Na to Li is still 0.39$e$, which is counterintuitive. One would expect instead 
neutral atoms for such a large separation.

\subsection{Character of the many-body wavefunction}

\begin{figure}
\begin{center}
\psfig{figure=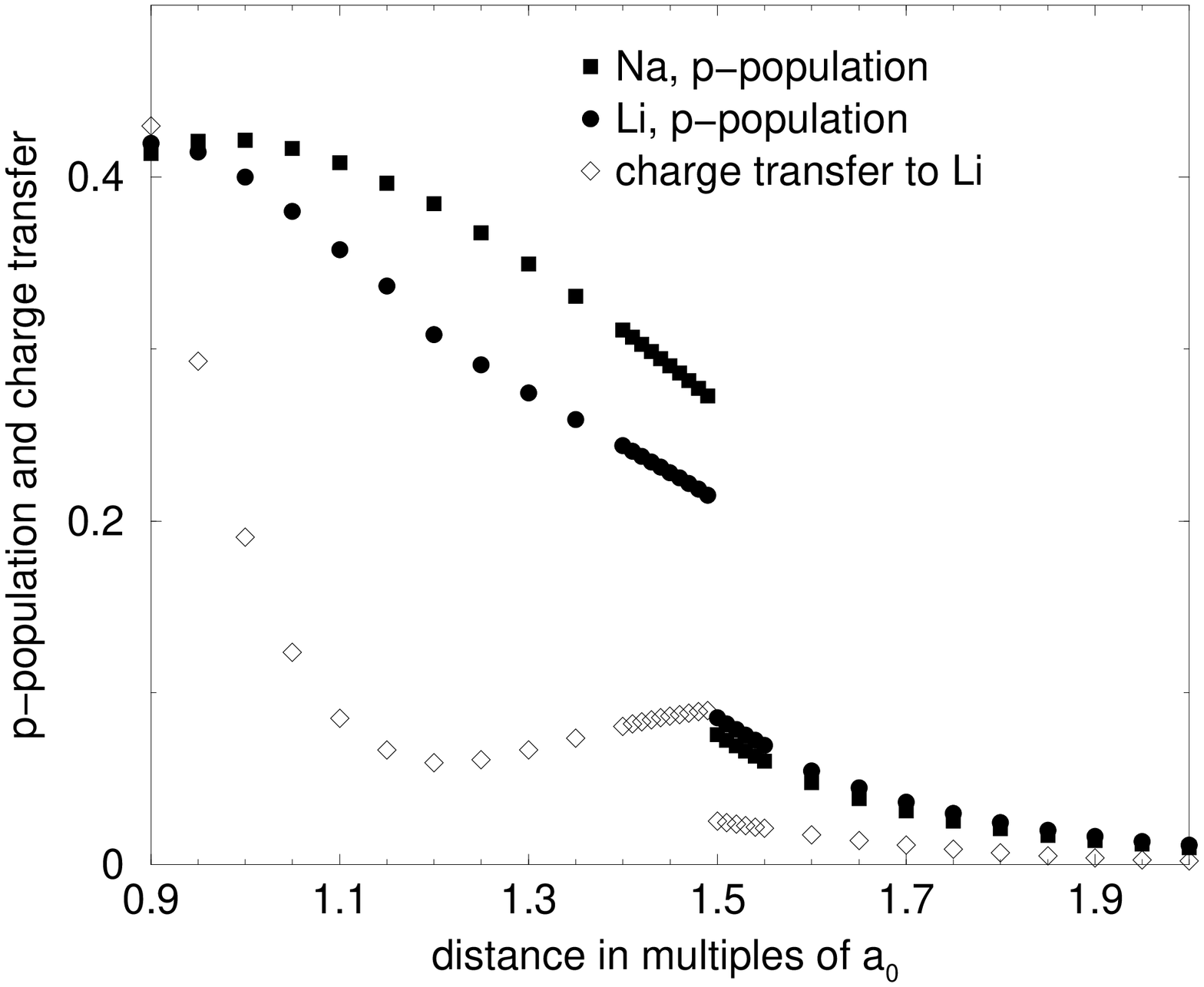,angle=-90,width=7cm}
\end{center}
\caption{\label{nalipop} The $p$ occupation of the Na atom and the Li atom in a Na$_5$Li$_5$ ring
versus the Li---Na distance. In addition the charge transfer from the Na atom to the Li atom is plotted.
The $p$ occupations and the charge transfer are calculated with the Mulliken 
population analysis with the MRACPF wavefunction.} 
\end{figure}

To describe the dissociation limit in the mixed system, i.e. the transition from a band insulator to the 
Mott insulator, a many-body treatment is necessary. 
We select as a finite system an equidistant Na$_5$Li$_5$ ring with $a_0$=5.954 a.u..
As for the Na$_{10}$ ring we perform a MCSCF calculation with 11 active orbitals, and on top 
of it a MRACPF calculation. 
Analysing the ground state energy for Na$_5$Li$_5$, we found as for pure Li or Na rings 
a kink in the MRACPF total energy curve for the  Na$_5$Li$_5$ ring.
Also here the quasi-degenerate $s$ and $p$ orbitals yield a change in
the character of the wavefunction. This is shown in more detail in the $p$-occupation
of the individual atoms (Fig.\ref{nalipop}). The $p$ population of the Na and Li atom
is high (greater than 0.2$e$) up to an interatomic distance of about 1.5$a_0$. It then jumps suddenly
to a value smaller than 0.1$e$. For the equilibrium interatomic distance (1.05$a_0$ for Na$_5$Li$_5$ ring
at MRACPF level) the $p$ population is about 0.4 for both Li and Na atoms.
As at the HF level, we found a charge transfer from Na to Li. At $a_0$ the correlation treatment
reduces the charge transfer by a factor of 2 compared to HF. Whereas at the HF level the
charge transfer remains nearly constant for all interatomic distances, the charge transfer
at the MRACPF level drops rapidly with increasing atomic distance, has a local minimum at about 1.2$a_0$ 
and then falls suddenly when the character of the wavefunction changes. For well separated atoms
we reach the expected neutral atom limit.\\
In summary, we can characterise the region of the band insulator as the region where we have
significant $p$-contribution in the many-body wavefunction, and the region of the Mott
insulator, where there is nearly no charge transfer and the $p$-contribution
in the wavefunctions is negligible.

\subsection{One-particle energy gap}

\begin{figure}
\begin{center}
\psfig{figure=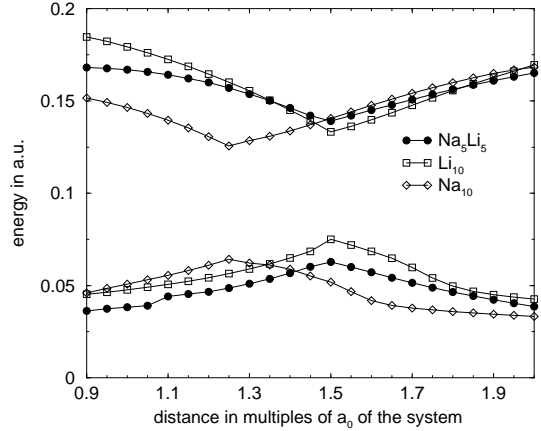,angle=-90,width=7cm}
\end{center}
\caption{\label{naligap} MRACPF values for the EA and IP
of the Na$_5$Li$_5$ ring, Na$_{10}$ and Li$_{10}$ ring
versus the interatomic distance. The scaling of the horizontal axis
always refers to the $a_0$ of the individual system.
}
\end{figure}

As for the pure Na ring, we calculate the energy gap of the Na$_5$Li$_5$ ring at the MRACPF level
due to adding and subtracting one electron to the system.
The IP and the EA of the Na$_5$Li$_5$ ring are plotted in Fig.\ref{naligap}.
For comparison the pure Na and Li data are shown, too. In the Na$_5$Li$_5$ ring the minimal gap
occurs at about 1.5$a_0^{\rm NaLi}$, as it does for the pure Li system whereas for Na the
minimal gap is at much smaller relative interatomic distances. The mixed system is dominated 
by the features of the pure Li system as there is a charge transfer from the Na to the Li atoms.
For the Mott insulator (larger interatomic distance than that, where the minimal gap occurs)
the pure and mixed system behave in the same manner, namely the IP and EA converge to the non-interacting atom
limit.
For the equilibrium distance the EA is almost the same for all three systems, whereas
the IP is the largest for Li and the smallest for Na. Comparing the NaLi with the Li system the difference of
the minimal gap to the gap at the equilibrium distance is not so pronounced.
Only for the "metallic" systems such as Li and Na is the closing of the gap faster than
in the band insulator NaLi. In other words, the electrons in the pure systems behave
more like free electrons than those in NaLi and feel stronger the effects of
the finite system than those in NaLi.
This difference is expected to be more pronounced when the rings get longer.

\subsection{Dipole Polarisability}

\begin{figure}
\begin{center}
\psfig{figure=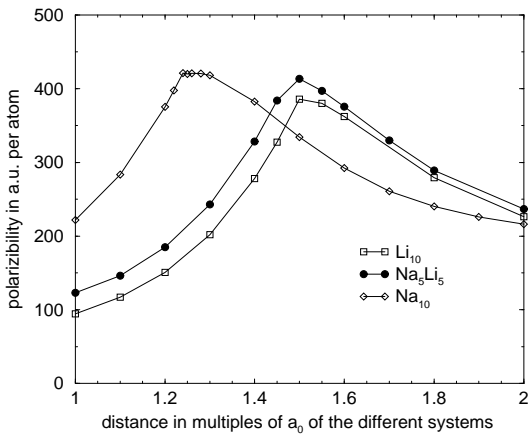,angle=-90,width=7cm}
\end{center}
\caption{\label{nalipol} The static dipole polarisability of the Na$_5$Li$_5$ ring, Na$_{10}$ and Li$_{10}$ ring
for the electric field
in the ring plane
 is plotted versus the interatomic distance for the MRACPF wavefunction.
The scaling of the horizontal axis
always refers to the $a_0$ of the individual system.}
\end{figure}

As a third quantity which can be used to indicate the MIT, 
we calculate the static electric dipole polarisability of the
system. 
For a metallic system the polarisability should be infinite, therefore a steep increase of the
polarisability can indicate an insulator-metal transition. The increase of the polarisation approaching the MIT
from the insulating side was also found in the Hubbard model\cite{aebischer01} named dielectric catastrophe.\\
The static electric dipole polarisability of the NaLi 
system is calculated by applying a static electric field of strength up to 0.003 a.u. for the ring in the ring plane.
We perform a quadratic fit with linear term for the energy of the system, 
$E({\mathcal{E}})=E(0)+\mu{\mathcal{E}}-\frac{1}{2}\alpha{\mathcal{E}}^2$, yielding the
polarisability $\alpha$ and the permanent dipole moment $\mu$. 
In Fig.\ref{nalipol} we plot the polarisability per atom for the 
Na$_5$Li$_5$ and the pure 
Na$_{10}$ and Li$_{10}$ rings.
The maximum of the polarisability occurs for all systems where the minimal gap occurs.
The mixed system does not behave differently from the pure systems. The only difference
for the mixed system is a small permanent dipole moment. For the 
equilibrium lattice constant it is about 0.07 a.u. and therefore only about $\frac{1}{10}$
of the dimer value with the same basis set. The permanent dipole moment vanishes slowly for 
larger interatomic distances.

\subsection{Lattice distortion}

Also for the mixed system we study the possibility of a lattice distortion in the system.
For interatomic distances in the Mott insulator regime the system dimerises
and forms Na---Li molecules.
In the region of the band insulator we found at the MRACPF level a dimerisation for a mean
interatomic distance larger than $a=1.3a_0$, below this point the 
equidistant arrangement of the atoms is the stable configuration.
$1.3a_0$ coincides with the charge transfer minimum in
the band insulator regime (see Fig.\ref{nalipop}). 
This is in agreement with results obtained from the ionic Hubbard model, where
in a limited region between two critical values of $\frac{U}{t}$ the bond alternation
is non zero\cite{fabrizio99,manmana03}.

\section{Conclusions}

We have investigated the analogue of the metal-insulator transition for one-dimensional
sodium and mixed lithium-sodium, applying high level quantum chemical ab-initio methods.
The MIT is modified from that of in the single-band Hubbard
model, when we have to deal with $s$ and $p$-orbitals per site.
At the transition point the character of the many-body wavefunction changes from significant $p$
to essential $s$ character.
Therefore it must be kept in mind that in a real solid the re-population of orbitals
belonging to different angular momenta may have similarly strong influence on the MIT as
changes in the ratio of the Hubbard $U$ to the kinetic energy.
We found that at approximately the same interatomic distance where the $p$ character
of the wavefunction changes so strongly, the one-particle
energy gap is minimal.\\
For the pure sodium system we have analysed the influence of the boundary conditions
by calculating a ring system (periodic boundaries) and a chain system (open boundaries).
Whereas in the ring system during the transition the character of the wavefunction changes rapidly 
from significant $p$
to pure $s$ character, the transition in the
chain system is smooth. Nevertheless a steep decrease of the $p$ contribution
is found for increasing interatomic distances.
The boundary conditions have no qualitative influence on the one-particle energy gap
as a function of the interatomic distance.
The position however where the minimal gap occurs, and the value of the minimal gap,
are strongly dependent on the boundary conditions. 
In the ring system a bond alternation only occurs above an interatomic distance
$a_{\rm Peierls}$ ($a_0\le a_{\rm Peierls} \le a_{\rm MIT}$) whereas the system with open boundaries
dimerises for all interatomic distances.\\
In the second part we have evaluated the same properties for the mixed lithium-sodium system.
As for pure systems, the character of the wavefunction changes from significant $p$
contribution to pure $s$ contribution. In addition the charge transfer from sodium to lithium
is nearly zero above the transition from a band to a Mott insulator.
The energy gap and the dipole polarisability are not much different from those of the pure systems.
This is probably due to the fact that finite sodium or lithium rings are not
real metals at the equilibrium distance, but behave more like band insulators with a
small band gap. Therefore the features we observe are due to different band gaps
in the systems rather than to a true qualitative change from a metal to an insulator.
As in the pure systems a bond alternation is found in a region below the
band to Mott insulator transition, but the dimerisation vanishes below 
an interatomic distance of 1.3$a_0$. This is in agreement with earlier results
for the ionic Hubbard model\cite{fabrizio99,manmana03}.\\
To analyse these properties in more detail and to approach the true metallic
regime, much longer systems have to be evaluated. Unfortunately this is not possible 
using straightforward
high-level quantum chemical methods. Some further approximations are neccesary, e.g. the application of
an incremental
scheme as was used for neutral lithium rings\cite{paulus03a}. 

\section*{Acknowledgements}

The authors would like to thank Prof. Peter Fulde (Dresden),
Prof. Krzysztof Rosciscewski (Krakow) and Prof. Hermann Stoll (Stuttgart)
for many valuable discussions.

\end{document}